\documentclass{article}
\usepackage[utf8]{inputenc}
\usepackage{arxiv,amsmath,amssymb,bm,bbm,mathtools,braket,hyperref,url,xcolor}

\usepackage[linesnumbered,ruled]{algorithm2e}

\SetCommentSty{mycommfont}

\usepackage{framed}
\renewcommand{\vec}{\bm}

\newcommand{\clonestate}{\coloneqq}
\newcommand{\applygate}{\leftarrow}

\usepackage{array}

\usepackage{multicol}

\SetNlSty{bfseries}{\color{gray}}{}

\title{Efficient classical calculation of the \\ Quantum Natural Gradient}
\author{
    Tyson Jones %\thanks{Corresponding author}
    \\
    Quantum Motion Technologies Ltd 
    \\
    Nexus, Discovery Way, Leeds, West Yorkshire LS2 3AA, United Kingdom
    \\
    \texttt{tyson.jones@quantummotion.tech} \\
}

\begin{document}
\maketitle

\begin{abstract}
    
    Quantum natural gradient has emerged as a superior minimisation technique in quantum variational algorithms~\cite{fubini_and_geometric,wierichs2020avoiding,van2020measurement}.
    Classically simulating the algorithm running on near-future quantum hardware is paramount in its study, as it is for all variational algorithms. In this case, state-vector simulation of the $P$-parameter/gate ansatz circuit does \textit{not} dominate the runtime; instead, calculation of the Fisher information matrix becomes the bottleneck, involving $\mathcal{O}(P^3)$ gate evaluations, though this can be reduced to $\mathcal{O}(P^2)$ gates by using $\mathcal{O}(P)$ temporary state-vectors. This is similar to the gradient calculation subroutine dominating the simulation of quantum gradient descent, which has attracted HPC strategies~\cite{broughton2020tensorflow,guerreschi2020intel} and bespoke simulation algorithms with asymptotic speedups~\cite{vari_grad_paper,luo2019yao}.
    We here present a novel simulation strategy to precisely calculate the quantum natural gradient in $\mathcal{O}(P^2)$ gates and $\mathcal{O}(1)$ state-vectors. While more complicated, our strategy is in the same spirit as that presented for gradients in Ref~\cite{vari_grad_paper}, and involves iteratively evaluating recurrent forms of the Fisher information matrix.
    Our strategy uses only ``apply gate", ``clone state" and ``inner product" operations which are present in practically all quantum computing simulators. It is furthermore compatible with parallelisation schemes, like hardware acceleration and distribution.
    %
    %While we do not present a similar scheme for noisy simulations via density matrices, our algorithm 
    %
    Since our scheme leverages a form of the Fisher information matrix for strictly unitary ansatz circuits, it cannot be simply extended to density matrix simulation of quantum natural gradient with non-unitary circuits~\cite{koczor2019quantum}.

\end{abstract}

% =====================
\section{Introduction}

Variational quantum eigensolving (VQE) is a promising application of first generation quantum computers~\cite{peruzzo2014variational,mcclean2016theory,preskill2018quantum}, with consequences for chemistry and material science~\cite{mcardle2020quantum,cao2019quantum}.
VQE aims to compute the minimum eigenvalue of some quantum operator by studying the states produced by a parameterized ansatz circuit.
A simple example is energy gradient descent, whereby the $P$ parameters, $\vec\theta$, are updated iteratively (with timestep $\Delta t$) under
\begin{align}
    \Delta \vec\theta =  - \Delta t \; \nabla \braket{E(\vec\theta)}.
\end{align}
The energy gradient 
\begin{align}
\nabla \braket{E(\vec\theta)} = 
\left\{
\frac{\partial \braket{E(\vec\theta)}}{\partial\theta_1}, \dots, 
\frac{\partial \braket{E(\vec\theta)}}{\partial\theta_P}
\right\},
\end{align}
is a length-$P$ vector which
can be tractably computed in a number of ways using a quantum device~\cite{schuld2019evaluating,li2017hybrid,mitarai2019methodology}.
Many superior methods to gradient descent have emerged, such as ADAM~\cite{kingma2014method}, BFGS~\cite{broyden1970convergence} and imaginary time evolution~\cite{mcardle2019variational}, though which require additional computation.
Recently, a quantum adaptation~\cite{stokes2020quantum} of natural gradient descent~\cite{amari1997neural} was shown to be superior still~\cite{wierichs2020avoiding}, and equivalent to imaginary time evolution in a noise-free setting~\cite{koczor2019quantum}.
The quantum natural gradient prescribes a change in parameters
\begin{align}
    g(\vec\theta) \, \Delta \vec\theta = - \Delta t \; \nabla \braket{E(\vec\theta)},
\end{align}
where in the absence of noise, $g(\vec\theta) = \Re[G(\vec\theta)]$ is the Fubini-Study metric tensor~\cite{stokes2020quantum}; a $P\times P$ matrix recently identified as the (classical) Fisher information matrix~\cite{koczor2019quantum,nat_grad_link_to_fubini}. We note quantum natural gradient was recently extended to make use of the \textit{quantum} Fisher information matrix to support non-unitary ans\"atze~\cite{koczor2019quantum}, though we do not explore this presently.
For a unitary ansatz $\hat U$ and ansatz state(vector) $\ket{\psi} = \hat U \ket{\text{in}}$, the Quantum Geometric Tensor $G(\vec\theta)$ is defined as
\begin{align}
    G_{ij}(\vec\theta) = \left\langle \frac{\partial \psi(\vec\theta)}{\partial \theta_i} ,
    \frac{\partial \psi(\vec\theta)}{\partial \theta_j}
    \right\rangle
    - 
    \left\langle \frac{\partial \psi(\vec\theta)}{\partial \theta_i} ,
     \psi(\vec\theta)
    \right\rangle
    \left\langle  \psi(\vec\theta), 
    \frac{\partial \psi(\vec\theta)}{\partial \theta_j}
    \right\rangle.
    \label{eq:geometric_quantum_tensor}
\end{align}
The quantum evaluation of this entire tensor~\cite{van2020measurement} requires $\mathcal{O}(P^2)$ rounds of sampling, each of which involve repeated but parallelisable evaluation of an ansatz circuit with $\mathcal{O}(P)$ gates. This means that if hypothetically an expected value could be calculated in one shot from an input state (like a classical simulation might), then the total number of quantum gates applied is $\mathcal{O}(P^3)$.
When the global phase of the ansatz state $\ket{\psi}$ is independent of the parameters, the latter terms of Eq~\ref{eq:geometric_quantum_tensor} vanish and $G(\vec\theta)$ becomes equivalent to the coefficient matrix in imaginary time evolution~\cite{van2020measurement,yuan2019theory}, and can leverage the slightly simpler circuits therein~\cite{mcardle2019variational}. These still prescribe a total of $\mathcal{O}(P^3)$ gates, before sampling.
We furthermore note there exist quantum circuits to more efficiently evaluate a block-diagonal approximation to the tensor $G(\vec\theta)$~\cite{stokes2020quantum}.

Like many variational algorithms which offer limited mathematical treatment, study of quantum natural gradient has been done numerically~\cite{koczor2019quantum,mcardle2019variational,wierichs2020avoiding}. However, the memory cost of classically representing a state-vector, and the time cost of simulating a gate, both grow exponentially with the number of qubits~\cite{poplavskiui1975thermodynamic}. 
There is a growing wealth of simulators which use hardware-acceleration and other high-performance computing techniques to parallelise simulation at the gate level~\cite{larose2019overview,fingerhuth2018open}, like QuEST~\cite{jones2019quest}, ProjectQ~\cite{steiger2018projectq}, qHIPSTER~\cite{smelyanskiy2016qhipster} and Quantum++~\cite{gheorghiu2018quantum}.
However, this is \textit{not} an optimal strategy for simulating variational algorithms, which can make use of parallelisation at a different granularity. For example, the ansatz circuit, for different values of the parameters, can be evaluated simultaneously in order to compute different elements of the gradient in parallel. Though this admits the same asymptotic costs, it can reduce overheads on high-performance hardware. This so-called ``batch strategy" is especially effective to deploy on distributed hardware, and has been recently integrated into Intel QS~\cite{guerreschi2020intel} and (non-distributed) Tensorfow Quantum~\cite{broughton2020tensorflow}.

For some variational algorithms, bespoke classical simulation routines can offer \textit{asymptotic} speedup. For example, the gradient vector in gradient descent was previously simulated with $\mathcal{O}(P^2)$ total gates, but was recently shown to admit an $\mathcal{O}(P)$ gate optimisation~\cite{vari_grad_paper,luo2019yao}. This negates the need for custom parallelisation routines, or the excessive memory to store multiple state-vectors simultaneously.
Such a strategy been implemented in Yao.jl~\cite{luo2019yao} and recently Qiskit~\cite{vari_grad_paper,cross2018ibm}.

In this manuscript, we present a novel algorithm to classically evaluate the Fisher information matrix (and hence the Fubini-Study metric), used in quantum natural gradient, in an asymptotically faster fashion than previous methods. Naively emulating the quantum evaluation and replacing sampling with direct inner products admits an $\mathcal{O}(P^3)$ runtime cost. In contrast, our scheme leverages a recurrent form of the tensor to evaluate it in $\mathcal{O}(P^2$) runtime, and a fixed memory overhead in terms of the number of parameters $P$.
This factor $P$ speedup enables the simulation of significantly deeper ansatz circuits, and hence the study of previously intractable systems. For example, even Ref~\cite{wierichs2020avoiding} which simulated natural gradient with a modest $40$ parameters could leverage our technique to simulate $\approx 250$ parameters in the same time.
We furthermore present an additional algorithm to simulate quantum natural gradient a factor $\approx3$ faster, though using $\mathcal{O}(P)$ additional state-vectors. This linearly growing memory cost may prove prohibitive in simulation of many qubits, or very deep ans\"atze.

\section{Derivation}
\label{sec:derivation}

For clarity, we illustrate recurrent forms of the individual terms of the geometric quantum tensor in Equation~\ref{eq:geometric_quantum_tensor}. We first consider the simpler expression, which is an inner product in the second term, which we'll notate as
\begin{align}
    T_i(\vec\theta) = \bigg\langle \psi(\vec\theta), \; \frac{\partial \psi(\vec\theta)}{\partial \theta_i}
    \bigg\rangle.
\end{align}

Assume $\psi(\vec\theta)$ is produced from an ansatz circuit $\hat{U}$ on some fixed input state $\ket{\text{in}}$, where $\hat{U}$ is composed of $P$ gates, $\hat{U}_i$, each with a unique parameter $\theta_i$. That is, 
\begin{align}
    \hat{U}(\vec\theta) &= \prod\limits_{k=P}^1 U_k(\theta_k) = \hat{U}_P(\theta_P) \dots \hat{U}_1(\theta_1),
&
&\text{and} &
    \ket{\psi(\vec\theta)} &= \hat{U}(\vec\theta) \ket{\text{in}}.
\end{align}
Then we have
\begin{align}
    T_i(\vec\theta) &= 
    \bigg\langle \text{in}\, \bigg|
    \left( \prod_{k=1}^{P} \hat U_k^\dagger
    \right)
    \left( \prod_{k=P}^{i+1} \hat U_k
    \right)
    \frac{\mathrm{d} \hat{U}_i}{\mathrm{d}\theta_i}
    \left( \prod_{k=i-1}^{1} \hat  U_k
    \right)
    \bigg|\, \text{in}  \bigg\rangle
    \\
    &=
    \bigg\langle \text{in}\, \bigg|
    \left( \prod_{k=1}^{i} \hat U_k^\dagger
    \right)
    \frac{\mathrm{d} \hat{U}_i}{\mathrm{d}\theta_i}
    \left( \prod_{k=i-1}^{1} \hat U_k
    \right)
    \bigg|\, \text{in}  \bigg\rangle,
\end{align}
where all gates $\{ \hat U_j : j > i \}$ have been eliminated.
For clarity, we define
\begin{align}
    \ket{\psi}_i &= \prod\limits_{k=i}^1 \hat{U}_k \ket{\text{in}},
    &
    &\implies \;\;\;
    \ket{\psi}_{i+1} = \hat{U}_{i+1} \ket{\psi}_i,
\end{align}
so that our term simplifies to
\begin{align}
T_i(\vec\theta) = \bra{\psi_i} \, \frac{\mathrm{d} \hat{U}_i}{\mathrm{d}\theta_i} \, \ket{\psi_{i-1}}.
\label{eq:t_recurrent_def}
\end{align}
This suggests a simple iterative strategy to calculate $T_i$ for all $i \in \{1, \dots, P\}$, in a total of $\mathcal{O}(P)$ gates, similar to that presented in Reference~\cite{vari_grad_paper}. However, we will instead integrate this calculation into the more sophisticated calculation of the other term below.

Now we consider only the first term in the geometric tensor of Equation~\ref{eq:geometric_quantum_tensor}, which we'll refer to as the Li tensor for its use in real-time simulation~\cite{li_benjamin_tensor};
\begin{align}
    L_{ij}(\vec\theta) = 
    \left\langle \frac{\partial \psi(\vec\theta)}{\partial \theta_i} ,
    \frac{\partial \psi(\vec\theta)}{\partial \theta_j}
    \right\rangle.
    \label{eq:L_def}
\end{align}
Then, 
\begin{align}
    L_{ij}(\vec\theta) = 
    \bigg\langle \text{in}\, \bigg|
    \left( \prod_{k=1}^{i-1} \hat U_k^\dagger
    \right)
    \frac{\mathrm{d} \hat U_i^\dagger}{\mathrm{d}\theta_i}
    \left( \prod_{k=i+1}^{P} \hat U_k^\dagger
    \right)
    \left( \prod_{k=P}^{j+1} \hat U_k
    \right)
    \frac{\mathrm{d} \hat U_j}{\mathrm{d}\theta_i}
    \left( \prod \limits_{k=j-1}^1 \hat{U}_k \right)
    \bigg|\, \text{in} \bigg\rangle.
\end{align}
Note $L_{ij} = {L_{ji}}^*$. For $i < j$, all gates after $j$ can be eliminated from the expression;
\begin{align}
    L_{ij}(\vec\theta)\Big|_{i<j} = 
    \bigg\langle \text{in}\, \bigg|
    \left( \prod_{k=1}^{i-1} \hat U_k^\dagger
    \right)
    \frac{\mathrm{d} \hat U_i^\dagger}{\mathrm{d}\theta_i}
    \left( \prod_{k=i+1}^{j} \hat U_k^\dagger
    \right)
    \frac{\mathrm{d} \hat U_j}{\mathrm{d}\theta_i}
    \left( \prod \limits_{k=j-1}^1 \hat{U}_k \right)
    \bigg|\, \text{in} \bigg\rangle.
\end{align}
Though in general not an L2 vector, we define $\ket{\phi}$ as
\begin{align}
    \ket{\phi}_{i,j}\big|_{i<j} &= 
    \left( 
        \prod\limits_{k=i}^j \hat{U}_k^\dagger
    \right)
    \frac{\mathrm{d}\hat{U}_j}{\mathrm{d}\theta_j}
    \ket{\psi}_{j-1}
    ,
\end{align}
and hence express the Li tensor as
\begin{align}
    L_{ij}(\vec\theta)\Big|_{i<j} 
    &=
    \big\langle \psi\,\big|_{i-1}
    \; 
    \frac{\mathrm{d} \hat U_i^\dagger}{\mathrm{d} \theta_i}
    \left( \prod_{k=i+1}^{j} \hat U_k^\dagger
    \right)
    \frac{\mathrm{d}\hat U_j}{\mathrm{d}\theta_j}
    \;
    \big|\,\psi\big\rangle_{j-1}
    \\
    &=
    \big\langle \psi\,\big|_{i-1}
    \; 
    \frac{\mathrm{d} \hat U_i^\dagger}{\mathrm{d} \theta_i} \,
    \big|\phi\big\rangle_{i+1,j}\;.
\end{align}
By observing the recurrent forms
\begin{align}
    \bra{\psi}_{i-1} &= \bra{\psi}_i \hat{U}_i,
    &
    &\text{and}
    &
    & \begin{gathered}
    \ket{\phi}_{i-1,j} = \hat U_{i-1}^\dagger \ket{\phi}_{i,j},
    \\
    \ket{\phi}_{j+1,j+1} = \frac{\mathrm{d}\hat{U}_{j+1}}{\mathrm{d}\theta_{j+1}} \ket{\psi}_{j},
    \end{gathered}
\end{align}
and for clarity, notating $\text{prod}\big[ \ket{a}, \ket b \big] = \braket{a|b}$, we make explicit a recurrency used by our strategy;
\begin{align}
    L_{ij}(\vec\theta)
    &=
    \text{prod}\left[ 
        \frac{\mathrm{d} \hat{U}_i}{\mathrm{d}\theta_i} \,
        \ket{\psi}_{i-1},
        \;\;
        \ket{\phi}_{i+1,j}
    \right],
    \label{eq:L_recurrent_def}
    \\
   \therefore \; L_{i-1,j}(\vec\theta)
    &=
    \text{prod}\left[ 
        \frac{\mathrm{d} \hat{U}_{i-1}}{\mathrm{d}\theta_{i-1}} \;
        \hat{U}_{i-1}^\dagger \,
        \ket{\psi}_{i-1},
        \;\;
        \hat{U}_i^\dagger \,
        \ket{\phi}_{i+1,j}
    \right].
\end{align}
This form reveals that given $\ket{\psi}_j$, we can produce $\ket{\phi}_{j+1,j+1}$ in one (derivative) gate operation, and from that iteratively produce $\ket{\phi}_{i,j}$ for every $i<j$, one gate operation at a time (similarly for $\ket{\psi}_i$), without any caching or creating additional memory. This means all $L_{ij}$ values, for $i<j$, can be computed in a total $\mathcal{O}(j)$ gate operations and $\mathcal{O}(1)$ space, in decreasing $i$ order. Since $\ket{\psi}_{j+1}$ is furthermore obtained from $\ket{\psi}_j$ by a single gate operation, all $L_{ij}$, for $i<j\le P$, may be calculated in $\mathcal{O}(P^2)$ gates.

This leaves the diagonal elements, which are
\begin{align}
    L_{ii}(\vec\theta) 
    &= \bigg\langle \text{in} \bigg|
    \left( 
        \prod\limits_{k=1}^{i-1} \hat{U}_k^\dagger
    \right)
    \frac{\mathrm{d}\hat{U}_i^\dagger}{\mathrm{d}\theta_i}
    \frac{\mathrm{d}\hat{U}_i}{\mathrm{d}\theta_i}
    \left( 
        \prod\limits_{k=i-1}^1 \hat{U}_k
    \right)
    \bigg|\text{in}\bigg\rangle
    \\
    &= \bra{\psi_{i-1}} \frac{\mathrm{d}\hat{U}_i^\dagger}{\mathrm{d}\theta_i}
    \frac{\mathrm{d}\hat{U}_i}{\mathrm{d}\theta_i} \ket{\psi_{i-1}}
    \\
    &= \braket{ \phi_{i,i} | \phi_{i,i} }
\end{align}
where although $\hat{U}^\dagger \hat{U} = \mathbbm{1}$, in general $\frac{\mathrm{d}\hat{U}}{\mathrm{d}\theta_i}$ is non-unitary, and hence $\braket{ \phi_{i,i} | \phi_{i,i} }$ is not necessarily unity.
Note that for typical rotation gates, the corresponding diagonal element need not be evaluated through simulation, but can be known \textit{a priori}. Any $N$-qubit unitary $\hat{V}$ is generated by
\begin{align}
    \hat{V}(\theta) &= \exp\left( \mathrm{i} \sum_{j=1}^{4^N} f_j(\theta) \hat{\sigma}_j \right),
    &
    \implies \;\;\;
    \frac{\mathrm{d}\hat{V}}{\mathrm{d} \theta}
    &=
    \mathrm{i} \sum\limits_{j=1}^{4^N} f_j'(\theta) \; \hat{\sigma}_j \; \hat{V}(\theta).
\end{align}
where $f_j: \mathbb{R} \mapsto \mathbb{R}$ and $\hat{\sigma}_j$ are Pauli strings. A ubiquitous example is a rotation around an axis of the Bloch sphere, like $R_X(\theta) = \exp(\mathrm{i} \theta/2 \hat{X})$; even multi-qubit rotation gates often admit a single term~\cite{li_benjamin_tensor}. Since $[\hat{A}, \exp(\hat A)] = \hat{0}$ and $\hat\sigma^\dagger=\hat\sigma$, our diagonal elements simplify to
\begin{align}
    L_{ii}(\vec\theta) &=
    \bra{\psi_{i-1}}
    \left( 
    - \mathrm{i} \sum\limits_{j=1}^{4^N} f_j'(\theta_i) \; \hat{\sigma}_j \; \hat{U}_i^\dagger(\theta)
    \right)
    \left(
     \hat{U}_i(\theta) \; \mathrm{i}
    \sum\limits_{k=1}^{4^N} f_k'(\theta_i) \; \hat{\sigma}_k 
    \right)
    \ket{\psi_{i-1}}
    \\
    &=
    \sum\limits_{j=1,k=1}^{4^N} f_j'(\theta_i) f_k'(\theta_i) \;
    \bra{\psi_{i-1}}
     \hat{\sigma}_j 
     \hat{\sigma}_k 
    \ket{\psi_{i-1}}
\end{align}
Whenever $\hat{U}_i(\theta_i)$ can be expressed as an exponential of a \textit{single} Pauli string with coefficient $f(\theta_i)$, then
\begin{align}
    L_{ii}(\vec\theta)
    &=
    f(\theta_i)^2 \braket{\psi_{i-1} | \sigma^2 | \psi_{i-1}} = f(\theta_i)^2.
    \label{eq:L_diag_when_single_pauli}
\end{align}

Hence $L_{ii}(\theta)$ can be calculated \textit{a priori} directly from the description of gate $\hat{U}_i(\theta_i)$. Often this quantity is independent of $\theta$, like for the common rotation gates;
\begin{align}
    \hat U_i(\theta_i) &= \exp\left(\mathrm{i} \,\theta\, \hat{\sigma}  / 2\right)
    &
    &\implies &
    f_i'(\theta_i) = \frac{1}{2}, & \;\;\; L_{ii}(\vec\theta) = \frac{1}{4}.
\end{align}
We point out that though that \textit{controlled} Pauli-string rotations lose this property;
\begin{gather}
    \hat{U}_i(\theta_i) 
    = |0\rangle\langle0| \otimes \mathbbm{1} + |1\rangle\langle 1| \otimes \exp(\mathrm{i} \, \theta \, \hat{\sigma} / 2),
    \;\;\;\;\;\;\;\;
    \implies \;\;\;\;\;\;\;\;
    \frac{\mathrm{d}\hat{U}_i^\dagger}{\mathrm{d}\theta_i}
    \frac{\mathrm{d}\hat{U}_i}{\mathrm{d}\theta_i}
    =
    \frac{1}{4} |1\rangle\langle 1| \otimes \mathbbm{1},
    \\
    \therefore 
    L_{ii}(\vec\theta) 
    =
    \frac{1}{4} 
    \langle \psi_{i-1}
    |1\rangle\langle 1|
    \psi_{i-1}\rangle
    = \frac{1}{4} | \braket{\psi_{i-1}|1} |^2.
\end{gather}

Having evaluated $L_{ij}$ and $T_i$ recurrently, the natural gradient is then simply
\begin{align}
    G_{ij}(\vec\theta) &= L_{ij}(\vec\theta) - {T_i}^*(\vec\theta) \; T_j(\vec\theta).
    \label{eq:g_t_L_def}
\end{align}

\section{Algorithm}
\label{sec:algorithm}

Our presented simulation strategy, Algorithm~\ref{alg:nat_grad_calc}, evaluates the Fisher information matrix (Eq.~\ref{eq:g_t_L_def}), using separate (but concurrent) evaluation of $L_{ij}$ and $T_{j}$ via the recurrent relations derived in Equations~\ref{eq:t_recurrent_def}~and~\ref{eq:L_recurrent_def} respectively. The presentation below is however slightly more complicated, since it re-uses some temporary state-vectors in order to minimise the number of gate operations needed (without compromising the constant memory overhead).

{\centering 
\begin{minipage}{\linewidth} % .7 \linewidth
\begin{algorithm}[H]
\DontPrintSemicolon
\SetKwInOut{Input}{Input}
\SetKwInOut{Output}{Output}

\Input{Initial state, which is input to the ansatz, $\ket{\text{in}}$}
\Input{Temporary state-vectors $\ket{\chi}$, $\ket{\psi}$, $\ket{\phi}$, $\ket{\lambda}$, $\ket{\mu}$}
\Input{Ansatz circuit $\prod_i \hat{U}_i(\theta_i)$}

\Output{Fisher information matrix $G_{ij} \, \forall \; 1 \le i,j \le P $}
\Output{$\ket{\psi}$ is left in ansatz state $\hat{U}_P \dots \hat{U}_1\ket{\text{in}}$}

\vspace{.1cm}
\tcp{Handle edge-cases}
%\vspace{.2cm}

$\ket{\chi} \clonestate \ket{\text{in}}$

$\ket{\chi} \applygate \hat{U}_1 $ % \ket{\chi} $
\tcp*{$\ket{\chi} = \hat{U}_1\ket{\text{in}}$, permanently}

$\ket{\psi} \clonestate \ket{\chi}$

$\ket{\phi} := \ket{\text{in}}$

$\ket{\phi} \applygate \frac{\mathrm{d}\hat{U}_1}{\mathrm{d}\theta_1} $ %\ket{\phi}$

$T_1 = \braket{\chi | \phi}$

$L_{1,1} = \braket{\phi | \phi}$

\vspace{.1cm}
\tcp{Compute $L_{i\le j}$ (Eq.~\ref{eq:L_recurrent_def}) and $T_k$ (Eq.~\ref{eq:t_recurrent_def})}
%\vspace{.2cm}

\For{$j \in \{2, \dots, P\}$}  {
    
    $\ket{\lambda} \clonestate  \ket{\psi} $
    \tcp*{$\ket{\lambda} =
        \hat{U}_{j-1} \dots \hat{U}_1 \ket{\text{in}}
        $}
    
    $ \ket{\phi} \clonestate \ket{\psi}$
    
    $ \ket{\phi} \applygate \frac{\mathrm{d} \hat{U}_j}{\mathrm{d}\theta_j} $ %\ket{\phi} $
    \tcp*{$\ket{\phi} = \frac{\mathrm{d} \hat{U}_j}{\mathrm{d}\theta_j} \hat{U}_{j-1} \dots \hat{U}_1 \ket{\text{in}}$}
    
    $L_{j,j} = \braket{\phi|\phi}$ 
    \tcp*{Skippable if $\hat{U}_j = \exp(\alpha \hat{\sigma})$ (Eq.~\ref{eq:L_diag_when_single_pauli})}
    
    \For{$i \in \{j-1, \, \dots\,, 1\}$}  {
    
        $\ket{\phi} \applygate \hat{U}_{i+1}^\dagger$ % \ket{\phi}$
        \tcp*{$\ket{\phi} =
        \hat{U}_{i+1}^\dagger \dots 
        \hat{U}_j^\dagger \frac{\mathrm{d} \hat{U}_j}{\mathrm{d}\theta_j} \hat{U}_{j-1} \dots \hat{U}_1 \ket{\text{in}}$}
        
        $\ket{\lambda} \applygate \hat{U}_i^\dagger $ %\ket{\lambda}$
        \tcp*{$\ket{\lambda} = \hat{U}_{i-1}\dots\hat{U}_1\ket{\text{in}}$}
        
        $\ket{\mu} \clonestate \ket{\lambda}$
        
        $\ket{\mu} \applygate \frac{\mathrm{d}\hat{U}_i}{\mathrm{d}\theta_i}
        $ %\ket{\mu}$
        \tcp*{$\ket{\mu} = \frac{\mathrm{d}\hat{U}_i}{\mathrm{d}\theta_i} \hat{U}_{i-1}\dots\hat{U}_1\ket{\text{in}}$}
        
        $L_{ij} = \braket{\mu | \phi}$
        
        \tcc*{$\braket{\mu|\phi} =
        \bra{\text{in}}
        \hat{U}_1^\dagger \dots \hat{U}_{i-1}^\dagger \frac{\mathrm{d}\hat{U}_i^\dagger}{\mathrm{d}\theta_i}
        \;\; \cdot \;\;
        \hat{U}_{i+1}^\dagger \dots \hat{U}_j^\dagger \frac{\mathrm{d} \hat{U}_j}{\mathrm{d}\theta_j} \hat{U}_{j-1} \dots \hat{U}_1 \ket{\text{in}}
        $}
    }
    
    $T_j = \braket{\chi|\phi}$
    \tcc*{$\braket{\chi|\phi} =
        \bra{\text{in}}
        \hat{U}_1^\dagger 
        \;\; \cdot \;\;
        \hat{U}_{2}^\dagger \dots \hat{U}_j^\dagger \frac{\mathrm{d} \hat{U}_j}{\mathrm{d}\theta_j} \hat{U}_{j-1} \dots \hat{U}_1 \ket{\text{in}}
        $}
    
    $\ket{\psi} \applygate \hat U_j $ % \ket{\psi}$
    \tcp*{$\ket{\psi} = \hat{U}_j \dots \hat{U}_1 \ket{\text{in}}$}
}

\vspace{.1cm}
\tcp{Unpack $T$ and $L$ values into the Hermitian quantum tensor $G$}
%\vspace{.2cm}

\For{$i \in \{1, \dots P \}$} {

    \For{$j \in \{1, \dots P\}$} {
        
        \uIf{$ i \le j $} {
            $G_{ij} = L_{ij} - {T_i}^* \, T_j$
        }
        \Else {
            $G_{ij} = {L_{ji}}^* - {T_i}^* \, T_j$
        }
    }
}

\caption{Calculating the complete Fisher information matrix $G_{ij}$ in a total $\mathcal{O}(P^2)$ gates / clone operations, and $\mathcal{O}(1)$ temporary state-vectors. Here, $\ket{a} \coloneqq \ket{b}$ denotes cloning state $\ket{b}$ into $\ket{a}$, while $\ket{a}\leftarrow \hat{U}$ denotes modifying $\ket{a}$ under the action of operator $\hat{U}$.
Comments on the right-hand-side indicate the state of the modified register after an operation.
}
\label{alg:nat_grad_calc}
\end{algorithm}
 \end{minipage}
 \par }

\section{Acknowledgements}

We thank Balint Koczor, Sam McArdle and Sam Jaques for helpful discussions.

\bibliographystyle{unsrt}
\bibliography{biblio.bib}

\begin{thebibliography}{10}

\bibitem{fubini_and_geometric}
James Stokes, Josh Izaac, Nathan Killoran, and Giuseppe Carleo.
\newblock Quantum natural gradient.
\newblock {\em Quantum}, 4:269, 2020.

\bibitem{wierichs2020avoiding}
David Wierichs, Christian Gogolin, and Michael Kastoryano.
\newblock Avoiding local minima in variational quantum eigensolvers with the
  natural gradient optimizer.
\newblock {\em arXiv:2004.14666}, 2020.

\bibitem{van2020measurement}
Barnaby van Straaten and B{\'a}lint Koczor.
\newblock Measurement cost of metric-aware variational quantum algorithms.
\newblock {\em arXiv preprint arXiv:2005.05172}, 2020.

\bibitem{broughton2020tensorflow}
Michael Broughton, Guillaume Verdon, Trevor McCourt, Antonio~J Martinez,
  Jae~Hyeon Yoo, Sergei~V Isakov, Philip Massey, Murphy~Yuezhen Niu, Ramin
  Halavati, Evan Peters, et~al.
\newblock Tensorflow quantum: A software framework for quantum machine
  learning.
\newblock {\em arXiv:2003.02989}, 2020.

\bibitem{guerreschi2020intel}
Gian~Giacomo Guerreschi, Justin Hogaboam, Fabio Baruffa, and Nicolas~PD Sawaya.
\newblock Intel quantum simulator: A cloud-ready high-performance simulator of
  quantum circuits.
\newblock {\em Quantum Science and Technology}, 5(3):034007, 2020.

\bibitem{vari_grad_paper}
Tyson Jones and Julien Gacon.
\newblock Efficient calculation of gradients in classical simulations of
  variational quantum algorithms, 2020.

\bibitem{luo2019yao}
Xiu-Zhe Luo, Jin-Guo Liu, Pan Zhang, and Lei Wang.
\newblock Yao. jl: Extensible, efficient framework for quantum algorithm
  design.
\newblock {\em arXiv preprint arXiv:1912.10877}, 2019.

\bibitem{koczor2019quantum}
B{\'a}lint Koczor and Simon~C Benjamin.
\newblock Quantum natural gradient generalised to non-unitary circuits.
\newblock {\em arXiv preprint arXiv:1912.08660}, 2019.

\bibitem{peruzzo2014variational}
Alberto Peruzzo, Jarrod McClean, Peter Shadbolt, Man-Hong Yung, Xiao-Qi Zhou,
  Peter~J Love, Al{\'a}n Aspuru-Guzik, and Jeremy~L O’brien.
\newblock A variational eigenvalue solver on a photonic quantum processor.
\newblock {\em Nature communications}, 5:4213, 2014.

\bibitem{mcclean2016theory}
Jarrod~R McClean, Jonathan Romero, Ryan Babbush, and Al{\'a}n Aspuru-Guzik.
\newblock The theory of variational hybrid quantum-classical algorithms.
\newblock {\em New Journal of Physics}, 18(2):023023, 2016.

\bibitem{preskill2018quantum}
John Preskill.
\newblock Quantum computing in the nisq era and beyond.
\newblock {\em Quantum}, 2:79, 2018.

\bibitem{mcardle2020quantum}
Sam McArdle, Suguru Endo, Alan Aspuru-Guzik, Simon~C Benjamin, and Xiao Yuan.
\newblock Quantum computational chemistry.
\newblock {\em Reviews of Modern Physics}, 92(1):015003, 2020.

\bibitem{cao2019quantum}
Yudong Cao, Jonathan Romero, Jonathan~P Olson, Matthias Degroote, Peter~D
  Johnson, M{\'a}ria Kieferov{\'a}, Ian~D Kivlichan, Tim Menke, Borja
  Peropadre, Nicolas~PD Sawaya, et~al.
\newblock Quantum chemistry in the age of quantum computing.
\newblock {\em Chemical reviews}, 119(19):10856--10915, 2019.

\bibitem{schuld2019evaluating}
Maria Schuld, Ville Bergholm, Christian Gogolin, Josh Izaac, and Nathan
  Killoran.
\newblock Evaluating analytic gradients on quantum hardware.
\newblock {\em Physical Review A}, 99(3):032331, 2019.

\bibitem{li2017hybrid}
Jun Li, Xiaodong Yang, Xinhua Peng, and Chang-Pu Sun.
\newblock Hybrid quantum-classical approach to quantum optimal control.
\newblock {\em Physical review letters}, 118(15):150503, 2017.

\bibitem{mitarai2019methodology}
Kosuke Mitarai and Keisuke Fujii.
\newblock Methodology for replacing indirect measurements with direct
  measurements.
\newblock {\em Physical Review Research}, 1(1):013006, 2019.

\bibitem{kingma2014method}
Diederik~P Kingma and Jimmy~Ba Adam.
\newblock {ADAM}: A method for stochastic optimization.
\newblock {\em arXiv:1412.6980}, 4, 2014.

\bibitem{broyden1970convergence}
Charles~George Broyden.
\newblock The convergence of a class of double-rank minimization algorithms 1.
  general considerations.
\newblock {\em IMA Journal of Applied Mathematics}, 6(1):76--90, 1970.

\bibitem{mcardle2019variational}
Sam McArdle, Tyson Jones, Suguru Endo, Ying Li, Simon~C Benjamin, and Xiao
  Yuan.
\newblock Variational ansatz-based quantum simulation of imaginary time
  evolution.
\newblock {\em npj Quantum Information}, 5(1):1--6, 2019.

\bibitem{stokes2020quantum}
James Stokes, Josh Izaac, Nathan Killoran, and Giuseppe Carleo.
\newblock Quantum natural gradient.
\newblock {\em Quantum}, 4:269, 2020.

\bibitem{amari1997neural}
Shun-ichi Amari.
\newblock Neural learning in structured parameter spaces-natural riemannian
  gradient.
\newblock In {\em Advances in neural information processing systems}, pages
  127--133, 1997.

\bibitem{nat_grad_link_to_fubini}
Naoki Yamamoto.
\newblock On the natural gradient for variational quantum eigensolver.
\newblock {\em arXiv preprint arXiv:1909.05074}, 2019.

\bibitem{yuan2019theory}
Xiao Yuan, Suguru Endo, Qi~Zhao, Ying Li, and Simon~C Benjamin.
\newblock Theory of variational quantum simulation.
\newblock {\em Quantum}, 3:191, 2019.

\bibitem{poplavskiui1975thermodynamic}
RP~Poplavski{\u\i}.
\newblock Thermodynamic models of information processes.
\newblock {\em Soviet Physics Uspekhi}, 18(3):222, 1975.

\bibitem{larose2019overview}
Ryan LaRose.
\newblock Overview and comparison of gate level quantum software platforms.
\newblock {\em Quantum}, 3:130, 2019.

\bibitem{fingerhuth2018open}
Mark Fingerhuth, Tom{\'a}{\v{s}} Babej, and Peter Wittek.
\newblock Open source software in quantum computing.
\newblock {\em PloS one}, 13(12):e0208561, 2018.

\bibitem{jones2019quest}
Tyson Jones, Anna Brown, Ian Bush, and Simon~C Benjamin.
\newblock {QuEST} and high performance simulation of quantum computers.
\newblock {\em Scientific reports}, 9(1):1--11, 2019.

\bibitem{steiger2018projectq}
Damian~S Steiger, Thomas H{\"a}ner, and Matthias Troyer.
\newblock Projectq: an open source software framework for quantum computing.
\newblock {\em Quantum}, 2:49, 2018.

\bibitem{smelyanskiy2016qhipster}
Mikhail Smelyanskiy, Nicolas~PD Sawaya, and Al{\'a}n Aspuru-Guzik.
\newblock {qHiPSTER}: The quantum high performance software testing
  environment.
\newblock {\em arXiv:1601.07195}, 2016.

\bibitem{gheorghiu2018quantum}
Vlad Gheorghiu.
\newblock Quantum++: A modern c++ quantum computing library.
\newblock {\em Plos one}, 13(12):e0208073, 2018.

\bibitem{cross2018ibm}
Andrew Cross.
\newblock The ibm q experience and qiskit open-source quantum computing
  software.
\newblock {\em APS}, 2018:L58--003, 2018.

\bibitem{li_benjamin_tensor}
Ying Li and Simon~C. Benjamin.
\newblock Efficient variational quantum simulator incorporating active error
  minimization.
\newblock {\em Phys. Rev. X}, 7:021050, Jun 2017.

\end{thebibliography}

\clearpage

\appendix

\section{Alternate algorithms}

We now demonstrate how a series of optimisations transform a naive, inefficient evaluation, into our presented algorithm. Besides clarifying the nature of our algorithm in a form more intuitive than the recurrent relations presented in Section~\ref{sec:derivation}, seeing each revised algorithm may help the reader identify their own implementation of evaluating the Fubini-Study matrix and the steps to optimise it.
Note in all these algorithms, gates are always applied directly onto a state, and products of gate matrices are never explicitly evaluated. This is because, while (e.g.) single-qubit gates operating on $N$-qubit state-vectors take time $\mathcal{O}(2^N)$, direct multiplication of gate matrices require `tensoring' these matrices to the full $2^N \times 2^N$ Hilbert space, and multiply at a cost $\mathcal{O}(2^{3N})$. There seems to be no practical regime where a multiplicative speedup of factor $P$ would outweigh the slowdown of $\mathcal{O}(2^{3N})$. So the primitives of these algorithms are strictly ``clone state'', ``apply gate to state'' and ``compute inner product``, each of which take time $\mathcal{O}(2^N)$, as in Ref.~\cite{vari_grad_paper}.

For simplicity, we describe only the evaluation of $L_{i\le j}$ (Equ.~\ref{eq:L_def}) and exclude the $T_i$ term, since computing it is a straightforward extension. The cost of these algorithms is collated in Table~\ref{tab:alg_abs_compare}.
To help explain the algorithms, we'll refer to the following sub-expressions of $L_{ij}$ as \textit{prefix}, \textit{infix} and \textit{suffix}; 
\begin{align}
    L_{ij}(\vec\theta)\Big|_{i<j} = 
    \underbrace{
        \bigg\langle \text{in}\, \bigg|
        \left( \prod_{k=1}^{i-1} \hat U_k^\dagger
        \right)
    }_{\textstyle \text{prefix}}
    \frac{\mathrm{d} \hat U_i^\dagger}{\mathrm{d}\theta_i}
    \underbrace{ 
        \left( \prod_{k=i+1}^{j} \hat U_k^\dagger
        \right)
    }_{\textstyle \text{infix}}
    \frac{\mathrm{d} \hat U_j}{\mathrm{d}\theta_i}
    \underbrace{
        \left( \prod \limits_{k=j-1}^1 \hat{U}_k \right)
        \bigg|\, \text{in} \bigg\rangle
    }_{\textstyle \text{suffix}}.
\end{align}

We begin with Alg.~\ref{alg:each_elem_super_dumb} which evaluates an unsimplified form of $L_{ij} \; \forall \; i, j < P$, to highlight the worst scaling.
For clarity, it generates $\ket{\frac{\partial \psi}{\partial \theta_i}}$ and $\ket{\frac{\partial \psi}{\partial \theta_j}}$ in separate statevectors $\ket{\phi_A}$ and $\ket{\phi_B}$ (respectively).
It unnecessarily visits every $(i,j)$ pair, recomputes $\ket{\phi_A}$ for every value of $j$, and simulates all gates $\hat{U}_{i+1}^\dagger \dots \hat{U}_P^\dagger \hat{U}_P \dots \hat{U}_{j+1}$ without cancellation. 

Alg.~\ref{alg:each_elem_sep_elim} is a simple improvement on Alg.~\ref{alg:each_elem_super_dumb}, calculating only $L_{i \le j}$ (since $L_{ij} = L_{ji}^*$), and cancelling the sub-expression $\hat{U}_{i+1}^\dagger \dots \hat{U}_P^\dagger \hat{U}_P \dots \hat{U}_{j+1}$. These changes come from the simplified analytic forms of $L_ij$, and are not in themselves algorithmic optimisations, nor achieve an asymptotic speedup.

The first algorithmic speedup is introduced in Alg.~\ref{alg:each_elem_sep_elim_cache_suffix}, where in lieu of preparing the state $\ket{\phi} = \frac{\mathrm{d}\hat{U}_j}{\mathrm{d}\theta_j} \hat U_{j-1} \dots \hat U_1 \ket{\text{in}}$ afresh every $j$ iteration, we instead compute the suffix $\hat{U}_{j-1} \dots \hat{U}_1 \ket{\text{in}}$ in a \textit{rolling} fashion, from the previous iteration's state of $\hat{U}_{j-2} \dots \hat{U}_1 \ket{\text{in}}$. We keep track of this suffix state in register $\ket{\psi}$, which is updated every $j$ iteration by a single gate. Still, each inner iteration of $i$ requires performing the full infix and prefix operations, which vary in number between $1$ and $j$, and our complexity remains $\mathcal{O}(P^3)$.

Alg.~\ref{alg:each_elem_sep_elim_cache_suffix_infix} additionally computes the \textit{infix} operators in a similar rolling fashion. The direction of the $i$ iteration is reversed (now visiting $i=j$ first, then decreasing) so that the state $\hat{U}_{i+1}^\dagger \dots \hat U_{j-1}^\dagger \frac{\mathrm{d}\hat{U}_j}{\mathrm{d}\theta_j} \hat U_{j-1} \dots \hat U_1 \ket{\text{in}}$ (infix + suffix) is obtained by applying one gate to the previous $i$ iteration's state, stored in $\ket{\lambda}$. However, each $i$ iteration must still compute the prefix state (with which to compute the inner product) in $i-1$ gates, maintaining the total $\mathcal{O}(P^3)$ scaling.

The final optimisation is to also evaluate the prefix state in a rolling fashion, as introduced in Alg.~\ref{alg:each_elem_sep_elim_cache_suffix_infix_prefix}. The trick is to identify that for a given $j$ iteration, the suffix state $\ket{\psi}$ has the form of the \textit{first} prefix state, needed by the first iteration $i=j$. Then, the subsequent prefix states can be obtained from the previous by a single gate, by `removing' a gate; that is, apply its adjoint. Alg.~\ref{alg:each_elem_sep_elim_cache_suffix_infix_prefix} stores the partial prefix state in $\ket{\mu}$. Now that every state in an $(i,j)$ iteration can be obtained from a previous iteration in a fixed number of gates, the total scaling becomes the number of unique $(i,j)$ elements to evaluate; $\mathcal{O}(P^2)$.
This makes  Alg.~\ref{alg:each_elem_sep_elim_cache_suffix_infix_prefix} a simplified form of the manuscript's main novel algorithm, Alg.~\ref{alg:nat_grad_calc}.

An entirely different but (factor) faster protocol is to independently evaluate all the states $\ket{\frac{\partial \psi}{\partial \theta_i}} \, \forall \; i \in \{1,\dots,P\}$, store each in an independent register and compute the inner products as a final step. This is done in Alg.~\ref{alg:each_deriv_sep}, and yields a $\mathcal{O}(P^2)$ runtime but an $\mathcal{O}(P)$ memory cost. Even this scheme can be sped-up (by an approx factor $2$) by evaluating the suffix state
$\hat{U}_{i-1} \dots \hat{U}_1 \ket{\text{in}}$ in a rolling fashion, as done in Alg.~\ref{alg:each_deriv_from_cache}. 
Note in these algorithms, since we're not computing inner products during iteration, we cannot rolling evaluate the expression ``from both ends" as before, since we do not know the final $\ket{\frac{\partial \psi}{\partial \theta_i}}$ state \textit{a priori}.
Hence further speedup is likely impossible.

\begin{figure}
    \centering
    \includegraphics[width=.64\textwidth]{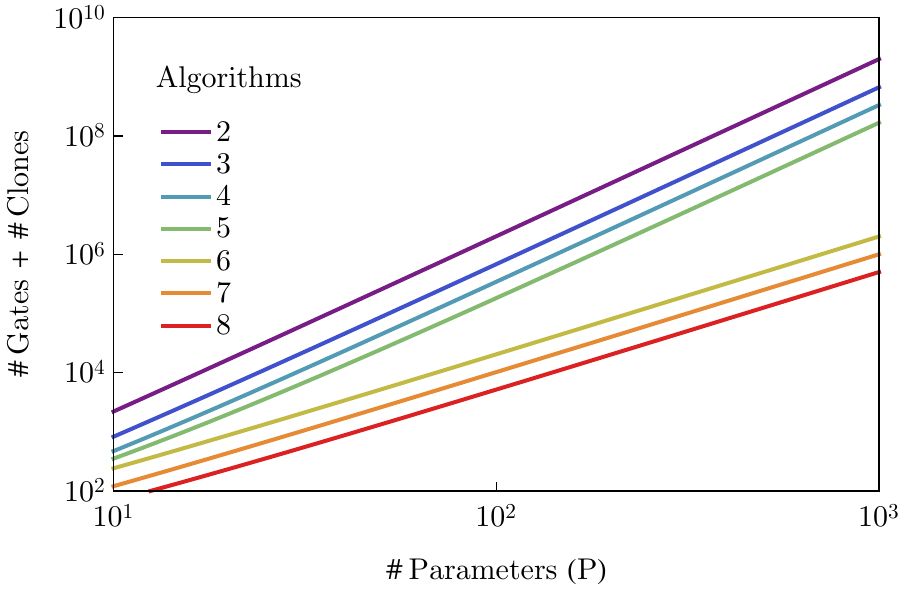}
    \caption{The total expected runtime cost (number of gates plus the number of clones) of Algorithms~\ref{alg:each_elem_super_dumb}-\ref{alg:each_deriv_from_cache}, as a function of the number of parameters.
    This plot visualises the expressions in Table~\ref{tab:alg_abs_compare}.
    }
    \label{fig:alg_costs_plot}
\end{figure}

The exact number of operations invoked by each algorithm, and the number of quantum registers used, is collated in Table~\ref{tab:alg_abs_compare}. The total number of operations is also visualised in Figure~\ref{fig:alg_costs_plot}, and makes clear the strict improvement for every practical regime (e.g. $P>10$) across Algorithms~\ref{alg:each_elem_super_dumb}-\ref{alg:each_deriv_from_cache}. For illustration, at $P = 1000$ parameters, Algorithms~\ref{alg:each_elem_sep_elim},~\ref{alg:each_elem_sep_elim_cache_suffix_infix_prefix} and~\ref{alg:each_deriv_from_cache} respectively cost $681750$, $20401$ and $5251$ total gates and clone operations respectively.

{\centering

    \makebox[\linewidth]{
    \begin{minipage}[t]{1.1\linewidth}

    \begin{minipage}[t]{.3\linewidth}
    \begin{algorithm}[H]
    \DontPrintSemicolon
    \SetKwInOut{Input}{Input}
    \SetKwInOut{Output}{Output}
    
    \Input{State-vector $\ket{\phi_A}$}
    \Input{State-vector $\ket{\phi_B}$}
    
    \For {$i \in \{1, \dots, P\}$} {
        
        \For {$j \in \{1, \dots, P\}$} {
        
            $\ket{\phi_A} \clonestate \ket{\text{in}} $
            
            $\ket{\phi_B} \clonestate \ket{\text{in}} $

            \For {$k \in \{1, \dots, i-1\}$} {
                $\ket{\phi_A} \applygate \hat{U}_k$
            }
            
            $\ket{\phi_A} \applygate \frac{\mathrm{d}\hat{U}_i}{\mathrm{d}\theta_i}$
            
            \For {$k \in \{i+1, \dots, P\}$} {
                $\ket{\phi_A} \applygate \hat{U}_k$
            }
            
            \For {$k \in \{1, \dots, j-1\}$} {
                $\ket{\phi_B} \applygate \hat{U}_k$
            }
            
            $\ket{\phi_B} \applygate \frac{\mathrm{d}\hat{U}_j}{\mathrm{d}\theta_j}$
            
            \For {$k \in \{j+1, \dots, P\}$} {
                $\ket{\phi_B} \applygate \hat{U}_k$
            }
            
            \tcp{$L_{ij} = \braket{\phi_A|\phi_B}$}
        }
    }
    
    \caption{Compute each \textit{element} independently, by preparing each \textit{state} independently.
    }
    \label{alg:each_elem_super_dumb}
    \end{algorithm}
    \end{minipage} \hfill
    \begin{minipage}[t]{.32\linewidth}
    \begin{algorithm}[H]
    \DontPrintSemicolon
    \SetKwInOut{Input}{Input}
    \SetKwInOut{Output}{Output}
    
    \Input{State-vector $\ket{\phi}$}
    
    \For {$j \in \{1, \dots, P\}$} {
        
        \For {$i \in \{1, \dots, j\}$} {
        
            $\ket{\phi} \clonestate \ket{\text{in}} $
            
            \For {$k \in \{1, \dots, j-1\}$} {
                $\ket{\phi} \applygate \hat{U}_k$
            }
            
            $\ket{\phi} \applygate \frac{\mathrm{d}\hat{U}_j}{\mathrm{d}\theta_j}$
            
            \For {$k \in \{j, \dots, i+1\}$} {
                $\ket{\phi} \applygate \hat{U}_k^\dagger$
            }
            
            $\ket{\phi} \applygate \frac{\mathrm{d}\hat{U}^\dagger_i}{\mathrm{d}\theta_i}$
            
            \For {$k \in \{i-1, \dots, 1\}$} {
                $\ket{\phi} \applygate \hat{U}_k^\dagger$
            }
            
            \tcp{$L_{ij} = \braket{\text{in}|\phi}$}
        }
    }
    \caption{Compute each \textit{element} independently, for $i \le j$ and eliminate gates $\{\hat{U}_{j+1}, \dots, \hat{U}_P\}$.
    }
    \label{alg:each_elem_sep_elim}
    \end{algorithm}
    \end{minipage} \hfill
    \begin{minipage}[t]{.32\linewidth}
    \begin{algorithm}[H]
    \DontPrintSemicolon
    \SetKwInOut{Input}{Input}
    \SetKwInOut{Output}{Output}
    
    \Input{State-vector $\ket{\psi}$}
    \Input{State-vector $\ket{\phi}$}
    \Input{State-vector $\ket{\lambda}$}
    
    $\ket{\psi} \clonestate \ket{\text{in}}$
    
    \For {$j \in \{1, \dots, P\}$} {
        
        $\ket{\phi} \clonestate \ket{\psi} $

        $\ket{\phi} \applygate \frac{\mathrm{d}\hat{U}_j}{\mathrm{d}\theta_j}$
            
        \For {$i \in \{1, \dots, j\}$} {
        
            $\ket{\lambda} \clonestate \ket{\phi}$
            
            \For {$k \in \{j, \dots, i+1\}$} {
                $\ket{\lambda} \applygate \hat{U}_k^\dagger$
            }
            
            $\ket{\lambda} \applygate \frac{\mathrm{d}\hat{U}^\dagger_i}{\mathrm{d}\theta_i}$
            
            \For {$k \in \{i-1, \dots, 1\}$} {
                $\ket{\lambda} \applygate \hat{U}_k^\dagger$
            }
            
            \tcp{$L_{ij} = \braket{\text{in}|\lambda}$}
        }
        
        $ \ket{\psi} \applygate \hat{U}_j $
    }
    \caption{Compute each \textit{element} independently, for $i \le j$, eliminate gates $\{\hat{U}_{j+1}, \dots, \hat{U}_P\}$, rolling cache suffix.
    }
    \label{alg:each_elem_sep_elim_cache_suffix}
    \end{algorithm}
    \end{minipage}

    \end{minipage} 
    } % END OF OVERFLOWING ROW

       \begin{minipage}[t]{.32\linewidth}
    \begin{algorithm}[H]
    \DontPrintSemicolon
    \SetKwInOut{Input}{Input}
    \SetKwInOut{Output}{Output}
    
    \Input{State-vector $\ket{\psi}$}
    \Input{State-vector $\ket{\phi}$}
    \Input{State-vector $\ket{\lambda}$}
    
    $\ket{\psi} \clonestate \ket{\text{in}}$
    
    \For {$j \in \{1, \dots, P\}$} {
        
        $\ket{\phi} \clonestate \ket{\psi} $

        $\ket{\phi} \applygate \frac{\mathrm{d}\hat{U}_j}{\mathrm{d}\theta_j}$
        
        \For {$i \in \{j, \dots, 1\}$} {
        
            $\ket{\lambda} \clonestate \ket{\phi}$
            
            $\ket{\lambda} \applygate \frac{\mathrm{d}\hat{U}^\dagger_i}{\mathrm{d}\theta_i}$
            
            \For {$k \in \{i-1, \dots, 1\}$} {
                $\ket{\lambda} \applygate \hat{U}_k^\dagger$
            }
            
            \tcp{$L_{ij} = \braket{\text{in}|\lambda}$}
            
            $\ket{\phi} \applygate \hat{U}_i^\dagger $
        }
        
        $ \ket{\psi} \applygate \hat{U}_j $
    }
    
    \caption{Compute each \textit{element} independently, for $i \le j$, eliminate gates $\{\hat{U}_j, \dots, \hat{U}_P\}$, rolling cache suffix and \textit{infix}.
    }
    \label{alg:each_elem_sep_elim_cache_suffix_infix}
    \end{algorithm}
    \end{minipage} \hspace{1cm}
    \begin{minipage}[t]{.32\linewidth}
    \begin{algorithm}[H]
    \DontPrintSemicolon
    \SetKwInOut{Input}{Input}
    \SetKwInOut{Output}{Output}
    
    \Input{State-vector $\ket{\psi}$}
    \Input{State-vector $\ket{\phi}$}
    \Input{State-vector $\ket{\lambda}$}
    \Input{State-vector $\ket{\mu}$}
    
     $\ket{\psi} \clonestate \ket{\text{in}}$
    
    \For {$j \in \{1, \dots, P\}$} {
        
        $\ket{\mu} \clonestate \ket{\psi}$
        
        $\ket{\phi} \clonestate \ket{\psi} $

        $\ket{\phi} \applygate \frac{\mathrm{d}\hat{U}_j}{\mathrm{d}\theta_j}$
        
        \For {$i \in \{j, \dots, 1\}$} {
        
            $\ket{\lambda} \clonestate \ket{\phi}$
            
            $\ket{\lambda} \applygate \frac{\mathrm{d}\hat{U}^\dagger_i}{\mathrm{d}\theta_i}$
            
            \tcp{$L_{ij} = \braket{\mu|\lambda}$}
            
            $\ket{\phi} \applygate \hat{U}_i^\dagger $
            
            \If{$i>1$} {
                $ \ket{\mu} \applygate \hat{U}_{i-1}^\dagger$
            }
        }
        
        $ \ket{\psi} \applygate \hat{U}_j $
    }
    
    \caption{Compute each \textit{element} independently, for $i \le j$, eliminate gates $\{\hat{U}_j, \dots, \hat{U}_P\}$, rolling cache suffix, infix and \textit{prefix}.
    }
    \label{alg:each_elem_sep_elim_cache_suffix_infix_prefix}
    \end{algorithm}
    \end{minipage} 
    
\par }

    % STATE CALC METHODS
    
\makebox[\linewidth]{
    
    \hfill \begin{minipage}[t]{.35\linewidth}
    \begin{algorithm}[H]
    \DontPrintSemicolon
    \SetKwInOut{Input}{Input}
    \SetKwInOut{Output}{Output}
    
    \Input{$P$ state-vectors $\{ \ket{\phi}_i \}$}
    
    \For {$i \in \{1, \dots, P\}$} {
        $\ket{\phi}_i \clonestate \ket{\text{in}} $
        
        \For {$k \in \{1, \dots, i-1\}$} {
            $\ket{\phi}_i \applygate \hat{U}_k$
        }
        
        $\ket{\phi}_i \applygate \frac{\mathrm{d}\hat{U}_i}{\mathrm{d}\theta_i}$
        
        \For {$k \in \{i+1, \dots, P\}$} {
            $\ket{\phi}_i \applygate \hat{U}_k$
        }
    }
    
    \tcp{ $L_{ij} = \braket{\phi_i|\phi_j} \;\; \forall \;\, i \le j$ }
    
    \caption{Compute each derivative \textit{state} independently.}
    \label{alg:each_deriv_sep}
    \end{algorithm}
     \end{minipage} \hfill
    \begin{minipage}[t]{.35\linewidth}
    \begin{algorithm}[H]
    \DontPrintSemicolon
    \SetKwInOut{Input}{Input}
    \SetKwInOut{Output}{Output}
    
    \Input{$P$ state-vectors $\{ \ket{\phi}_i \}$}
    \Input{State-vector $\ket{\psi}$}
    
    $\ket{\psi} \clonestate \ket{\text{in}}$
    
    \For {$i \in \{1, \dots, P\}$} {

        $\ket{\phi}_i \clonestate \ket{\psi}$
        
        $\ket{\phi}_i \applygate \frac{\mathrm{d}\hat{U}_i}{\mathrm{d}\theta_i}$
        
        \For {$k \in \{i+1, \dots, P\}$} {
            $\ket{\phi}_i \applygate \hat{U}_k$
        }
        
        $\ket{\psi} \applygate \hat{U}_i$
    }
    
    \tcp{ $L_{ij} = \braket{\phi_i|\phi_j} \;\; \forall \;\, i \le j$ }
    
    \caption{Compute each derivative \textit{state}, rolling cache suffix.}
    \label{alg:each_deriv_from_cache}
    \end{algorithm}
     \end{minipage}  \hfill
}

\begin{table}
    \setlength\extrarowheight{6pt}
    \centering
    \begin{tabular}{|c|cr|c|c|cr|}
        \hline
         Algorithm & \# Registers & & \# Gates & \# Clones & \# Gates + \# Clones & \\ \hline
         \ref{alg:each_elem_super_dumb}
         & $2$ & $=\mathcal{O}(1)$ & $2 P^3$ & $2P^2$
         & $2P^3 + 2P^2$
         & $ = \mathcal{O}(P^3)$
         \\ \hline
         \ref{alg:each_elem_sep_elim} & $1$ & $=\mathcal{O}(1)$
         & 
         $\frac{2 }{3}P^3 + P^2+\frac{1}{3}P$
         &
         $ \frac{1}{2}P^2+\frac{1}{2}P $
         &
         $\frac{2}{3}P^3 + \frac{3}{2}P^2 + \frac{5}{6} P$
         & $ = \mathcal{O}(P^3)$
         \\ \hline
         \ref{alg:each_elem_sep_elim_cache_suffix}
         & $3$ & $=\mathcal{O}(1)$ & 
         $ \frac{1}{3}P^3+\frac{1}{2}P^2+\frac{13}{6}P $
         & 
         $ \frac{1}{2}P^2+\frac{3 }{2}P + 1 $
         &
         $\frac{1}{3} P^3 + P^2 + \frac{11}{3} P + 1$
         & $ = \mathcal{O}(P^3)$
         \\ \hline
         \ref{alg:each_elem_sep_elim_cache_suffix_infix}
         & 
         $3$ & $=\mathcal{O}(1)$ & $\frac{1}{6}P^3+P^2+\frac{11 }{6}P$
         &
         $ \frac{1}{2}P^2+\frac{3 }{2}P+1 $
         &
         $\frac{1}{6}P^3 + \frac{3}{2}P^2 + \frac{10}{3}P + 1$
         & $ = \mathcal{O}(P^3)$
         \\ \hline 
         \ref{alg:each_elem_sep_elim_cache_suffix_infix_prefix}
         &
         $4$ & $=\mathcal{O}(1)$
         &
         $ \frac{3}{2} P^2 +\frac{3 }{2}P $
         &
         $\frac{1}{2}P^2+\frac{5 }{2}P+1$
         &
         $ 2 P^2 + 4 P + 1$
         & $ = \mathcal{O}(P^2)$
         \\ \hline 
         \ref{alg:each_deriv_sep} & 
         $P$ & $=\mathcal{O}(P)$ & 
         $P^2 + P$ & $P$ 
         &
         $ P^2 + 2 P$
         & $ = \mathcal{O}(P^2)$
         \\ \hline 
         \ref{alg:each_deriv_from_cache} & 
         $P+1$ & $=\mathcal{O}(P)$
         & $\frac{1}{2} P^2 + \frac{3}{2}P$ & $P+1$
         &
         $ \frac{1}{2}P^2 + \frac{5}{2}P + 1$
         & $ = \mathcal{O}(P^2)$
         \\ \hline
    \end{tabular}
    \vspace{.2cm}
    \caption{Runtime and memory costs of Algorithms~\ref{alg:each_elem_super_dumb}-\ref{alg:each_deriv_from_cache}, which compute $L_{ij}$ (Equ.~\ref{eq:L_def}).
    Algorithms~\ref{alg:each_elem_super_dumb}-\ref{alg:each_elem_sep_elim_cache_suffix_infix_prefix} have fixed memory costs, and include progressive optimisations to shrink the runtime from $\mathcal{O}(P^3)$ to $\mathcal{O}(P^2)$;
    indeed Alg.~\ref{alg:each_elem_sep_elim_cache_suffix_infix_prefix} is a simplified form of our main presented Alg.~\ref{alg:nat_grad_calc}.
    Algorithms~\ref{alg:each_deriv_sep} and \ref{alg:each_deriv_from_cache} are alternate strategies with superior runtime performances, but growing (and possibly prohibitive) memory costs.
    All algorithms ultimately compute a total of $(P^2+P)/2$ inner products (being the number of upper triangular elements in $[L_{ij}]$, including the diagonal), with the exception of Alg.~\ref{alg:each_elem_super_dumb} which needlessly computes the full $P^2$ matrix for illustration purposes.
    }
    \label{tab:alg_abs_compare}
\end{table}

\end{document}